\documentclass[preprint,review, 12pt,authoryear]{elsarticle}

\journal{Journal of Theoretical Biology}

\usepackage{natbib}
\usepackage{amsfonts}           
\def\BE{\mathbb{E}}             
\def\BP{\mathbb{P}}             
\usepackage{graphicx}           
\usepackage{setspace, amsthm}
\usepackage{graphicx}
\usepackage{amsbsy,amssymb,amsmath, amsthm}
\usepackage{booktabs}
\usepackage{psfrag}
\usepackage{rotating}
\usepackage{enumerate}
\usepackage{lineno}
\usepackage{tabularx}
\usepackage{multirow}
\usepackage{fancyhdr}

\def\pgf[#1][#2][#3][#4]{\pi(#1_{0:#2} |#3_{0:#4},\theta, \psi )}

\def\Var{\mathbb{V}ar}

\def\bn {{\bf n}}
\def\bm {{\bf m}}

\def\ignore#1{{}}

\title{
Detecting non-binomial sex allocation when developmental mortality operates\\
\vspace{1cm}
RUNNING TITLE: Detecting non-binomiality
} 

\author[1]{Richard D. Wilkinson}
\author[2,3]{Apostolos Kapranas}
\author[2]{Ian C.W. Hardy}

\date{}

\address[1]{Corresponding author: School of Mathematics and Statistics, University of Sheffield, Sheffield, S3 7RH, UK. {\it r.d.wilkinson@sheffield.ac.uk}}
\address[2]{School of Biosciences, University of Nottingham, Sutton Bonington Campus, Loughborough, LE12 5RD, UK}
\address[3]{Laboratory for Fundamental and Applied Research in Chemical Ecology, University of Neuchâtel, Neuchâtel, 2000 Switzerland}

\usepackage[margin=1in]{geometry}

\begin{document}


\maketitle



\section*{Abstract}

Optimal sex allocation theory is one of the most intricately developed areas of evolutionary ecology. Under a range of conditions, particularly under population sub-division, selection favours sex being allocated to offspring non-randomly,  generating non-binomial variances of offspring group sex ratios. 
Detecting non-binomial sex allocation is complicated by stochastic  developmental mortality, as offspring sex can often only be identified on maturity with the sex of non-maturing offspring remaining unknown.  We show that current approaches for detecting non-binomiality  have limited ability to detect non-binomial sex allocation when developmental mortality has occurred.
We present a new procedure using an explicit model of sex allocation and mortality and develop a Bayesian model selection approach  (available as an R package). We use the double and multiplicative binomial distributions to model over- and under-dispersed sex allocation and show how to calculate Bayes factors for comparing these alternative models to the null hypothesis of binomial sex allocation.
The  ability  to detect non-binomial sex allocation is greatly increased, particularly in cases where mortality is common. The use of Bayesian methods allows for the quantification of the  evidence in favour of each hypothesis, and our modelling approach provides an improved descriptive capability over existing approaches. 
We use a simulation study to demonstrate  substantial improvements in power for detecting non-binomial sex allocation in situations  where current methods fail, and we illustrate the approach in real scenarios using empirically obtained datasets on the sexual composition of groups of gregarious parasitoid wasps.

{\bf Key words:} Sex ratio; under-dispersion;   Bayes factor; Markov chain Monte Carlo

\section{Introduction}

The null model of sex allocation theory is the D\"{u}shing-Fisher theory of equal investment \citep{West2009}. When populations are both large and have unbiased sex ratios, selection for variance in the sexual composition of offspring groups is predicted to be absent \citep{Kolman1960}. Under these conditions mothers will not be selectively penalized if they randomly allocate sex to offspring, with fixed probability of $p= 0.5$ that the offspring is male, independently of the sex of previous offspring. Thus, the number of males in each offspring group would have binomial variance, i.e., $np(1-p)$, where $n$ is the number of offspring. In smaller populations and under sex ratio bias ($p\not = 0.5$), stabilizing selection for low sex ratio variance is predicted, i.e., variance less than $np(1-p)$ \citep{Verner1965, West2009}. Selection on sex ratio variance is likely to be strong when populations are structured into small reproductive subgroups within which offspring mate with each other on maturity and prior to the dispersal of the daughters \citep[local mate competition;][]{Hamilton1967}; here, selection favours the evolution of low sex ratio variance, especially when one or a very few mothers contribute offspring to the locally mating group \citep{Green_etal1982, Hardy1992, NagelkerkeHardy94, Nagelkerke1996, West_etal1998}. This is because low variance maximizes the production of mated daughters, a close correlate of maternal fitness. If one male is sufficient to mate successfully with all females within a group and all offspring in the group are progeny of one mother, then the optimal sexual composition is one male and the remainder of the group being females \citep{Green_etal1982}. Similar arguments predict low variance under local resource competition (a generalization of local mate competition) and its converse, local resource enhancement \citep{Lambin1994}.
Variance in the number of males among groups lower than expected under binomial sex allocation is known as under-dispersion, and sex allocation is then termed {\it precise} \citep{Green_etal1982, Lambin1994, Nagelkerke1996}.

Control of sex allocation can be detected in some organisms by direct observation of sexually differential aspects of individual offspring production, such as maternal movements during egg laying, or the placement of offspring, or by non-random production sequences \citep{Cole1981, Hardy1992, Heinsohn_etal1997, Krackow_etal02, Khidr_etal2013, Ambrosini2014} but such evidence is not often available. 
Empiricists must more frequently rely on the statistical analysis of offspring group sex ratios to detect whether sex allocation is being controlled or whether it is, for instance, binomial, as might be the null-expectation under several chromosomal mechanisms of sex-determination \citep{Aviles2000, Krackow_etal02, Ewen_etal2003a, Macdonald_etal2008, Postma_etal2011}. Furthermore, empirical evaluations of sex ratio variance can provide tests of explicit predictions of sex ratio theory \citep[e.g.,][]{Lambin1994, Morgan1994, Hardy1995, Hardy_etal1998, NagelkirkeSabelis98, West_etal1998, Kapranas_etal2011, Khidr_etal2013, Bowers_etal2013}.

One practical problem often faced by investigations of sex ratios and sex ratio variance is that information on the sexual compositions of offspring is available at maturity but not at the time of sex allocation, and it is not uncommon for some offspring to die before maturity, \citep[e.g.,][]{Hardy_etal1998, Dyrcz_etal2004, Ewen_etal2004,Forsyth_etal2004, Dietrich-Bischoff_etal2006,Oigarden_etal2013}.
Provided it has a stochastic component, developmental mortality will act to increase the variance of observed sex ratios, making initially under-dispersed data appear closer to binomial. This effect is expected on logical grounds (Section \ref{sect:Current}) and has been shown empirically both within and across several species of organisms with group structured mating (\citealt{Hardy_etal1998, Kapranas_etal2011, Khidr_etal2013}; see also \citealt{Dyrcz_etal2004} and \citet{Dietrich-Bischoff_etal2006}). Current statistical approaches to assessing sex ratio variance \citep{Krackow_etal02} are, however, based on the implicit assumption that developmental mortality does not operate, and they consequently lack power to detect non-binomiality, unless mortality rates are low.

Our aim  is to show that by introducing a model that  represents the biological processes that generated the data (sex allocation followed by mortality), we can substantially improve  our ability to detect underlying biological behaviours. We also demonstrate the advantage of using more descriptive statistical approaches such as estimating effect sizes (with measures of confidence), rather than relying on null-hypothesis significance testing, where the small dataset sizes mean we often fail to clear an arbitrary significance hurdle (usually $\alpha=0.05$) even when the data indicate  phenomena of interest.
 We begin by evaluating the performance, under developmental mortality, of the statistical methods commonly used to detect non-binomial sex ratio variance. We find that the power of these methods is adversely affected by developmental mortality. We then develop an alternative 
 approach that explicitly models the mortality process. This has much improved power for detecting non-binomial sex allocation, particularly when there is high mortality or datasets are small. 

\section{Terms and notation}\label{sect:Notation}

We define some terms and notation before describing current approaches and their limitations, and then introduce our new approach for detecting non-binomial sex allocation. A summary of the notation is provided in Table \ref{table:Notation}. The methods  developed are general, but are likely to most readily be applied to egg-laying organisms such as birds, parasitoid wasps, fig wasps and phytoseiid mites \citep{Hardy1992, NagelkirkeSabelis98, West_etal1998, West2009, Bowers_etal2013}, and this is reflected in the terminology we adopt \citep[for a mammalian example see][]{Macdonald_etal2008}. 
 Assume that we have a dataset containing data on $C$ different clutches of eggs, all of which were laid in comparable environmental conditions. Offspring group size is called {\it clutch size} at the time of production (egg-laying)  and  {\it brood size} at the time of offspring maturity: brood size is less than clutch size when developmental mortality occurs.

A {\it primary} dataset consists of counts of the number of eggs and their sex for each clutch. Let $N_i$ denote the number of eggs laid in the i$^{th}$ clutch, and $M_i$ be the number of those $N_i$ eggs that are male. A primary dataset is  the collection $\{(N_i, M_i)\}_{i=1}^C$. However, for most empirical investigations $M_i$ is not observed, as the sex of an offspring cannot be easily determined from the eggs: it is usual to wait until the eggs hatch and develop to the point at which offspring sex can be discriminated  \citep[e.g.,][]{Dietrich-Bischoff_etal2006, Khidr_etal2013}. It is also usual that a proportion of the eggs fail to mature, due to some form of developmental mortality, and consequently their sex cannot be recorded.

A {\it secondary} dataset consists of counts of $n_i$, the number of offspring that reach maturity (brood size) and $m_i$, the number of those offspring that are male, with the complete secondary dataset denoted $\{(n_i, m_i)\}_{i=1}^C$. Although a small number of experiments have been conducted where primary datasets are obtained, either directly from genetic characteristics of eggs \citep{Dijkstra1985,  Hardy_etal1998, NagelkirkeSabelis98, Khidr_etal2013} or through selective statistical procedures \citep{Dyrcz_etal2004, Kapranas_etal2011}, the vast majority of analyses have been conducted using secondary datasets \citep[e.g.,][]{Hardy1992, West_etal1998, NagelkirkeSabelis98, Mackauer_etal2002, Dietrich-Bischoff_etal2006,  Kapranas_etal2008}.

Our 
null hypothesis about sex allocation, $H_0$, is that there is a sex ratio $p$ (the proportion of offspring that are male), and that each egg is male with probability $p$ independently of all other eggs in the clutch, i.e., that the distribution of sex ratios is binomial 
\begin{equation}\label{eqn:binomial}
M_i \sim \operatorname{Bin}(N_i, p).
\end{equation}
 The alternative hypothesis, $H_1$, is that the number of males is non-binomially distributed, that is, either over- or under-dispersed when compared to the binomial distribution. Note that these are hypotheses  about primary sex ratios, not secondary sex ratios.

\begin{table}
\centering
  \begin{tabular}{l l}
\toprule
  Symbol& Definition\\
  \midrule
$C$ & Number of clutches in the dataset\\
$N$ & Number of eggs laid (primary)\\
$M$ & Number of eggs laid that are male (primary)\\
$n$ & Number of offspring that reach maturity (secondary)\\
$m$ & Number of males that reach maturity  (secondary)\\
$D$ & The complete observed dataset, i.e., $D=\{(n_i, m_i)\}_{i=1}^C$\\
\midrule
$p$ & Sex ratio$\dagger$ (proportion of eggs that are male)\\
$\psi$ & Dispersion parameter\\
$\lambda$ & Average clutch size\\
$d$ & Mortality rate\\
\midrule
$H_0, \;H_1$ & Null and alternative hypotheses\\
$U$ & Test statistic for the Meelis'  test\\
$R$ & Descriptive ratio contrasting observed and expected variance\\
$s^2$ & McCullagh's dispersion estimator\\
$\mathcal{S}$ & Clutch sizes observed in the data, i.e., $\{k: n_j=k \mbox{ for some } j\}$\\
$v_k$ & Number of clutches of size $k$, i.e., $\sum_{i=1}^C \mathbb{I}_{n_i=k}$\\
$s_k^2$& Empirical variance of the number of males in clutches of size $k$\\
$B_{01}$ &  Bayes factor for comparing $H_0$ with $H_1$\\
\bottomrule
\end{tabular}
\caption{Summary of notation used in this article. Letters in {\bf bold} font indicate vector quantities,  indices (e.g., $n_i$) indicate an instance of that variable, and hats (e.g., $\hat{p}$) indicate estimates. $\dagger$Care needs to be taken with interpretation of $p$ in the multiplicative binomial model as $p$ is no longer the expected sex ratio when $\psi\not = 0$.}
\label{table:Notation}
\end{table}

\section{Current approaches for detecting non-binomial sex allocation}\label{sect:Current}

Several methods have been used for the statistical analysis of sex ratio variances \citep{James1975, Green_etal1982, NagelkirkeSabelis98, West_etal1998, Krackow_etal02}. Whilst these methods can work well when applied to primary sex ratio data, this is not usually available, and so these methods are instead applied to secondary data, effectively treating them as if they were primary data.
 Not considering or ignoring that mortality has occurred thus violates  the assumptions behind each approach;  this results in  a lack of statistical power, often  leading to incorrect conclusions.

The first method for detecting departures from the binomial distribution, is a formal statistical test derived by E. Meelis \citep{NagelkirkeSabelis98}, which we  refer to as the {\it Meelis test} \citep{Krackow_etal02}. The test is  a comparison of the estimated variance with the variance under the assumption of a binomial distribution, and is derived by calculating the distribution (under the null hypothesis) of $\sum m_i^2$ conditional on $\sum m_i$. 
A test statistic $U$ (see supplementary material for details) is defined which can be shown to have a standard normal distribution under $H_0$, provided $C$ is sufficiently large. Large negative values of $U$ indicate under-dispersion, and large positive values over-dispersion;   typically, the test is applied by calculating the p-value $\BP(|U|>|u_{obs}|)$, where $\BP$ denotes probability, with small values taken to indicate departure from the null hypothesis.

There are several difficulties with applying the Meelis test to the datasets used in empirical studies of sex-allocation.
Firstly,  the test assumes that the binomial random variables are observed directly, which is not the case when using secondary data (using $m_i$ instead of $M_i$). 
Secondly,  the test is derived for use on random variables from a binomial distribution with fixed size ($n_i=n$ for all $i$), whereas for real data, the values of $n_i$  vary between broods, with  datasets typically consisting of a range of brood sizes. It is common practice to collect all the broods of a certain size (e.g., all $m_i$ such that $n_i=j$), then calculate the $U$-statistic, denoted $U_j$ for those broods, before  combining them using 
\[U = \frac{\sum U_j}{\sqrt{ |\mathcal{S}|}}\]
to give a single statistic $U$, where $\mathcal{S}=\{k: n_j=k \mbox{ for some } j\}$ is the collection of clutch sizes observed in the dataset.
If each $U_j \sim N(0,1)$, then $U \sim N(0,1)$. However, the Meelis test was derived for large sample sizes. In practice, there may only be a small number of clutches with $n_i=j$, and so each $U_j$ may not be well approximated by a standard normal distribution and hence, $U$ may not have a $N(0, 1)$ distribution either.

James' test \citep{James1975} is an alternative to the Meelis test that is often used for analysing datasets containing small clutches of unequal sizes. It involves calculation of a test statistic \citep[][give details]{Krackow_etal02}, which is known to be approximately normally distributed under the assumption of binomial sex ratios (no mortality). Large positive values  indicate over-dispersion, and negative values under-dispersion. It is known to be  less powerful for a single clutch size than the Meelis test (and suffers from the same difficulties as the Meelis test), but is included in our analysis  for completeness. 

The descriptive ratio $R$ is also used:
\begin{linenomath}
$$ R = \frac{ \sum_{k \in \mathcal{S}} v_k s_k^2 }{\sum_{k \in \mathcal{S}}v_k k \hat{p}_k (1-\hat{p}_k) }$$
\end{linenomath}
where $s_k^2$ is the empirical variance of the number of males in clutches of size $k$, i.e., $s_k^2 = \Var(\{m_i: n_i=k\})$, and $v_k=\sum_{i=1}^C \mathbb{I}_{n_i=k}$ is the number of clutches which have size $k$. The denominator is the sum of the variances if assuming a binomial distribution, where $\hat{p}_k$ is the estimated sex ratio for clutches of size $k$, i.e.,
\begin{linenomath}
$$\hat{p}_k = \frac{1}{kv_k}\sum_{i=1}^C m_i \mathbb{I}_{n_i=k}.$$
\end{linenomath}
The rationale for using $R$, is that it is the observed variance of the number of males, divided by the variance that would occur if the number of males was binomially distributed \citep{Krackow_etal02}. We expect to observe $R\approx 1$  if the data are binomially distributed, with  $R<1$ for under-dispersed data. 
  \citet{McCullaghNelder1989} introduce a further estimator of dispersion, which is a sum of ratios rather than a ratio of sums 
  \begin{linenomath}
  $$s^2 = \frac{1}{C-1} \sum_{i=1}^C  \frac{(m_i - \hat{p} n_i)^2}{n_i\hat{p}(1-\hat{p})}\quad
\mbox{where}\quad \hat{p} = \frac{\sum{m_i}}{\sum n_i},$$ 
\end{linenomath}
and should be    interpreted in the same way as the  $R$ statistic.

The effect of mortality is to make the data appear less under-dispersed (more binomial), as  mortality has the effect of increasing the variance of the number of males. To see this, imagine a species which has perfect precision, with each mother laying the same number of male and female eggs every time, so that  the sex ratio variance is zero. Stochastic mortality would introduce an element of randomness to the sexual composition of the offspring groups, such that  secondary datasets  may even resemble binomial random variables under sufficiently high rates of  mortality (see  Section \ref{sect:Cflorus}).

\subsection{Evaluation of current approaches when developmental mortality occurs}

To illustrate the limitations of current approaches, we simulate synthetic under-dispersed primary datasets, and then simulate the mortality process to produce synthetic secondary datasets. By applying the approaches described above, and repeating the process numerous times, we can examine their performance under varying levels of mortality.

We simulated sample experimental datasets as follows: for $i=1, \ldots, C$,
\begin{enumerate}

\item Simulate the clutch size from a Poisson distribution: $N_i \sim Po(\lambda)$, where $\lambda$ is the average clutch size.

\item Simulate the number of males in the $i^{th}$ clutch, $M_i$, from an under-dispersed multiplicative binomial distribution (Section \ref{sect:Bayes}).

\item Simulate  the secondary dataset by assuming each of the $N_i$ eggs has probability $d$ of not reaching maturity, and count the number of females and males that survive.

\end{enumerate}


\noindent We used a primary dataset on the parasitoid wasp {\it Goniozus legneri} \citep{Khidr_etal2013}, a species known to produce a strongly under-dispersed primary sex ratio, to estimate parameter values for the synthetic data model, and used these estimates fixed throughout the simulation study ($\lambda=10.0, p= 0.00278$, and $\psi= 0.445$, where $p$ and $\psi$ are parameters in the multiplicative binomial distribution,  which is an under-dispersed distribution - see Section \ref{sect:secondary}). We varied the size of the simulated experiment $C$, and the mortality rate $d$, and for each pair of values we simulated 10,000 synthetic datasets, and averaged the test statistics found across the  replicates. This allows the effectiveness of all the  procedures to be examined across a range of dataset sizes,  $C$, and  mortality rates $d$.

\begin{figure}
 \centering
  \includegraphics[width=\textwidth]{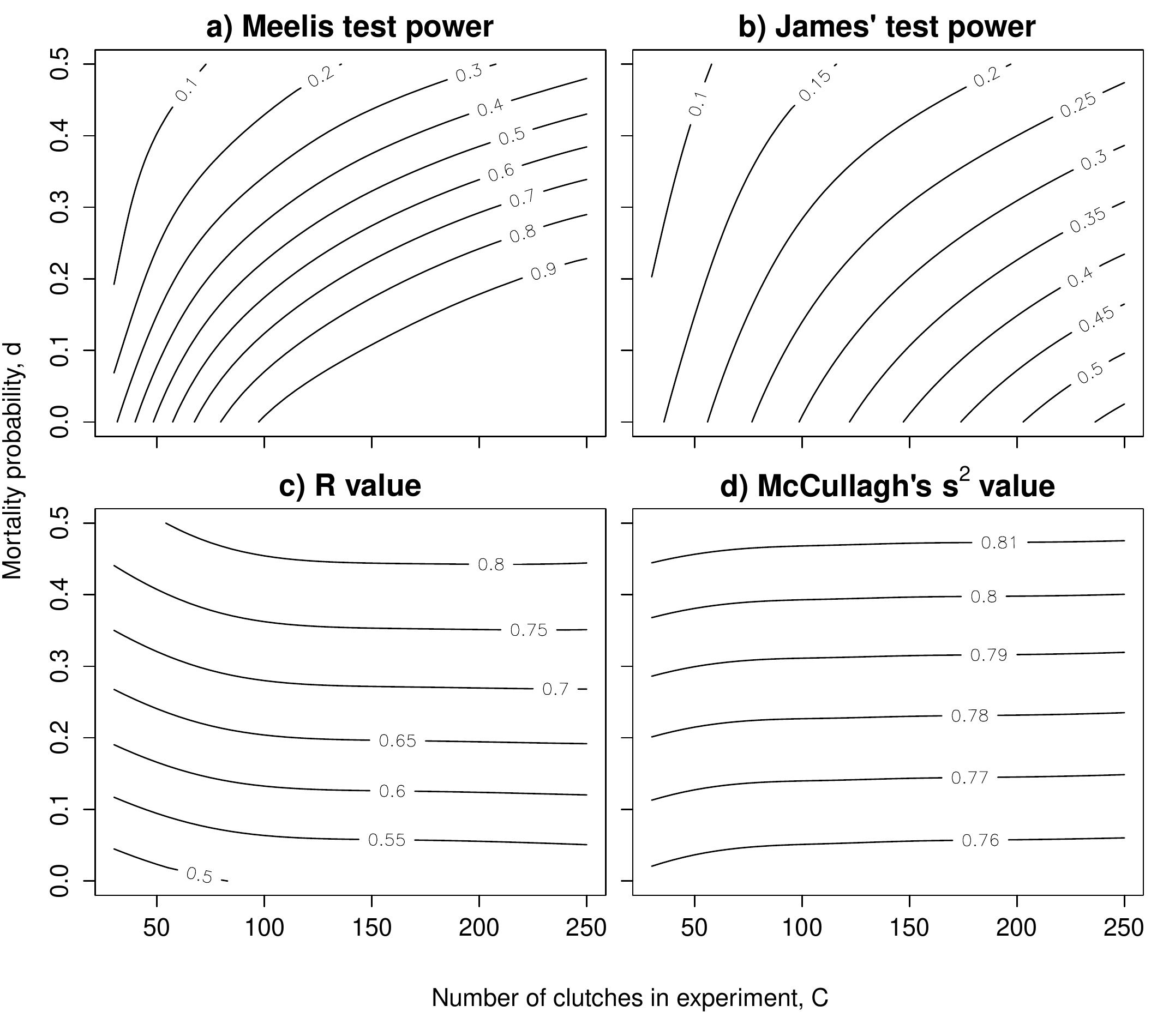}
\caption{a) and b) show contour plots of the power of the two-sided Meelis and James tests; c) and d) are contour plots of the values of descriptive statistics $R$ and McCullagh's $s^2$, all as a function of $C$ and $d$. The values were estimated using 10,000 randomly generated datasets, using parameter values estimated from  data on {\it G. legneri} primary sex ratios.}
\label{fig:Meelis1}
\end{figure}

The performance of  a hypothesis test can be measured by its power for a given significance level, where power is the probability of detecting non-binomial sex allocation when it occurs (i.e., power =  $1-\BP(\mbox{Type II error}) = \BP(\mbox{reject } H_0 \mid H_1)$). Contour plots of the power of the Meelis and James tests (at significance level 0.05) as a function of the number of clutches in the dataset and the mortality rate  show  that the  test lacks power if the number of clutches used is small or if the mortality rate is moderate-to-large (Fig. \ref{fig:Meelis1}a,b). For example, for a dataset containing 50 clutches with a mortality rate of 10\% there is only a 35\% probability of correctly detecting under-dispersion. The power of the test is lower still if lesser degrees of under-dispersion are assumed as it becomes harder to detect (we used reasonably large under-dispersion of $\psi=0.445$).



Fig. \ref{fig:Meelis1}c    shows the effect of mortality on $R$. The expected value of $R$ increases towards 1 as the mortality rate increases, so that species with a high mortality rate will have $R$ values consistent with binomial sex allocation, even if their primary sex ratios are under-dispersed. 
Fig. \ref{fig:Meelis1}d shows the same information for McCullagh's $s^2$. This can be seen to be less affected by mortality and so its use should be preferred to $R$.  The number of clutches in the experiment  has only a minor effect  on the expected value of both statistics. However,  it strongly affects the variance of the estimate (not shown), and for smaller experiments the observed values can vary widely, and so without appropriate confidence intervals for both statistics, 
they have little value.

There are two (related) reasons for the lack of power  in these approaches. The first is that mortality  increases the variance of the secondary values ($n_i, m_i$) compared to the primary values ($N_i, M_i$) making under-dispersion harder to detect. The second is that the tests do not take into account the fact that mortality has occurred, and consequently the additional variance is incorrectly interpreted as being consistent with binomial sex ratios.

\section{A new test for detecting non-binomial sex allocation} \label{sect:Bayes}

By explicitly modelling mortality we  develop a test that has improved statistical power as well as an increased descriptive capability. Our null hypothesis is a binomial model of sex allocation,  which we compare to two different  generalisations of the binomial distribution, the   multiplicative binomial and the double binomial distributions, both of which can model  over- and under-dispersion. Our model for the data then consists of a mortality model applied to the output of the sex allocation model. We use Bayesian model selection to determine which model is best supported by the data. The more intricate computational details are given in the supplementary material; here we focus on the broad outline of the approach.


\subsection{A model of secondary data}\label{sect:secondary}

We assume we have data on $C$ different broods from comparable environmental conditions, so that they can be considered to be statistically exchangeable.  Note that the unobserved primary counts $N_i$ and $M_i$, and the corresponding secondary values after mortality has occurred, $n_i$ and $m_i$, must satisfy the inequalities 
\begin{equation}\label{eqn:inequalities}
 N_i \geq n_i,\quad M_i \geq m_i \,\,\mbox{  and  }\,\,  N_i - n_i \geq M_i - m_i.
\end{equation}
We consider three models for the data, which differ only in the distribution of the sex allocation, i.e., the distribution of $M_i$ given $N_i$. The first  is the binomial model, with $M_i | N_i, p \sim \operatorname{Bin}(N_i,p)$,
which corresponds to the null hypothesis in Section 2. The second  is the multiplicative binomial distribution introduced by \citet{Altham78}:
\begin{equation} \label{eqn:multbinom}
 \BP(M | N, p, \psi) = c(p, \psi) \binom{N}{M} p^M (1-p)^{N-M} {\rm  e}^{\psi M(N-M)},
\end{equation}
where $c(p, \psi)$ is an intractable normalising constant. The two parameters are a probability $p$, and a dispersion parameter $\psi$. The third, introduced by \citet{Efron86}, is the double binomial model
\begin{equation}\label{eqn:doublebinom}
\BP(M | N, p, \psi) = c(p, \psi) \binom{N}{M}\frac{ N^{N\psi}p^{M(\psi+1)} (1-p)^{(N-M)(\psi+1)}}{ M^{M\psi} (N-M)^{(N-M)\psi}}
\end{equation}
where $c(p, \psi)$ is again an intractable normalising constant.
Note that when $\psi=0$, both the multiplicative and double binomial distributions reduce to the binomial distribution. These models are the key part of our procedure, corresponding to the  alternative hypothesis in Section \ref{sect:Notation}, as they both model the three cases of interest:

\begin{enumerate}[(i)]
 \item binomial sex allocation when $\psi=0$
\item under-dispersed sex allocation when $\psi >0$
\item over-dispersed sex allocation when $\psi < 0$.
\end{enumerate}
Unfortunately neither of these two distributions arises from a simple physical mechanism. Familarity does allow an intuition to develop about the meaning of $\psi$, but our usage here does not require any interpretation beyond that given above, and that larger values of $\psi$ indicate more under-dispersion than small values etc.
Care also needs to be taken with interpretation of $p$, as the expected value of $M$ is no longer $Np$ for the multiplicative binomial distribution, except when $\psi=0$, and so $p$ can no longer be considered to be the sex ratio (the expected sex ratio, $\BE\left(\frac{M}{N}\right)$, can be determined by Monte Carlo integration). 
We include both models as alternatives, as different datasets fit different models better, and this makes the detection of under-dispersion more likely.

We use the same model of mortality in each hypothesis and assume that each egg has probability $d$ of dying before maturity, and thus of not being counted in the secondary dataset, independently of its sex and the other eggs in the clutch, i.e., we assume mortality is  binomially distributed:
\begin{equation} \label{eqn:mort1} 
n_i | N_i, d \sim \operatorname{Bin}(N_i, d).
\end{equation}
The distribution of $m_i$ can then be shown, by a label permuting argument, to be
\begin{equation}\label{eqn:mort2}
 \BP(m | M, N, n, d) = \frac{\binom{M}{M-m} \binom{N-M}{ N-n-M+m}}{\binom{N}{N-m}}.
\end{equation}
 

We use two complimentary approaches for detecting departures from binomial sex allocation, the first based on estimation of effect size, and the second on hypothesis testing \citep{Nakagawa_etal2007}. The simpler approach is to estimate the effect size, measured by the dispersion parameter $\psi$,  by finding its posterior distribution $\pi(\psi | D)$. This parameter indicates whether sex allocation is binomial, over-, or under- dispersed, as well as how strong the effect is.
Posterior credibility intervals for $\psi$ can be used to assess the precision of the estimates and indicate informally
 whether  the data are consistent with $H_0$ ($\psi=0$). We describe methodology to do this below, the code is provided in the {\tt precision} R package, and applications are described in Section \ref{sect:results}. 

While various authors recommend estimation over hypothesis testing \citep{Robert2001, Gelman_etal2003, Nakagawa_etal2007}, relying solely on estimation of $\psi$ does not always provide the clarity required. For example, if the posterior distribution contains some support for $\psi=0$, but the posterior mode is not close to $0$, it can be difficult to judge whether or not data are under-dispersed using only the posterior distributions (Section \ref{sect:Gthail}). 
Instead, we wish to obtain the probability that sex allocation is under-dispersed, i.e., the posterior probability that each of the three models $M_0$, $M_1$ and $M_2$ are true conditional upon the data: $\BP(M_0 | D)$, $\BP(M_1| D)$,  and $\BP(M_2 | D)$.
These probabilities only make sense in a  Bayesian setting, although  note that $p$-values obtained from classical hypothesis tests, such as the Meelis test, are often incorrectly interpreted in this way \citep{Goodman08}.
 
Bayesian model selection requires calculation of the Bayes factor \citep{Jeffreys39, KassRaftery95}, which is 
defined as the ratio of the evidence for two different hypotheses (or models)
\begin{equation}\label{eqn:BF}B_{01}=\frac{\pi(D|H_1)}{\pi(D|H_0)}.\end{equation}
Values of $B_{01}$ greater than 1 indicate evidence in favour of $H_1$ (over $H_0$) and values less than 1 indicate evidence for $H_0$ (over $H_1$). 
\citet{Jeffreys39} suggested interpretation of the strength of evidence in favour of a hypothesis according to the magnitude of the Bayes factor is shown in Table \ref{table:BF}. 
The Bayes factor $B_{ij} = \BP(D|H_j)/\BP(D|H_i)$  can also be interpreted by noting that it is  the ratio between the posterior and prior odds in favour of $H_j$ over $H_i$
 \[\frac{\BP(H_j|D)}{\BP(H_i|D)} = B_{ij} \frac{\pi_j}{\pi_i} \]
where $\pi_j$ is the prior probability of $H_j$. Table \ref{table:BF}   contains the posterior probabilities of $H_1$ being true for various Bayes factor ranges  when we assume the hypotheses are equally likely {\it a priori}.


\begin{table}
\centering
  \begin{tabular}{lll}
\toprule
  $B_{01}$ range & $\BP(H_1|D)$ range&  Interpretation\\
\midrule
1--3 & 0.5-0.75 &  Barely worth mentioning\\
3--10 & 0.75 - 0.91    &Substantial\\
10--30 & 0.91-0.97 & Strong\\
30--100 & 0.97- 0.99& Very strong\\
$>100$ & 0.99-1 & Decisive\\
\bottomrule
 \end{tabular}
\caption{Jeffreys' suggested  interpretation of the Bayes factor for strength of evidence in favour of $H_1$  over $H_0$. Values of $1/B_{01}=B_{10}$ give the strength of evidence for $H_0$ over $H_1$. Also shown are  the corresponding ranges of the posterior probability for $H_1$ given the data, in the case where we assign equal prior probability to both hypotheses.}
\label{table:BF}
\end{table}

Bayes factors provide a powerful alternative to frequentist hypothesis tests, and have several advantages over classical methods. The first is that they provide a way to evaluate the evidence  in favour of a hypothesis, in contrast to the classical approach which only rejects or accepts the null hypothesis for a particular error rate. 
This is particularly useful in datasets where the effect size or the sample size are small, or where the mortality rate is high, as we can quantify the strength of the evidence for under-dispersion in the data, even if there is not enough evidence to formally reject the null hypothesis. For instance,  for analysis of data on {\it Goniozus thailandensis} (Section \ref{sect:Gthail}),  the Meelis test finds $p >0.05$ and thus concludes that there is no evidence to reject the null hypothesis, whereas the Bayesian approach  reports that the posterior probability of the double binomial model being the true model is 0.79, with the probability of the binomial model being true only 0.14.
When combined with the  posterior distribution of $\psi$, which is concentrated on values greater than 0, this  strongly suggests that this species produces under-dispersed sex ratios, a message that is lost if we only report the decision from the Meelis test.

\subsection{Parameter estimation}

We now describe how to find the posterior distribution of the parameters $\theta = (\psi, p, d, \lambda)$ given the data $D=\{(n_i, m_i)\}_{i=1}^C$, which we denote $\pi(\theta |D)$. This distribution represents our beliefs about the parameters after training the model to take the observed experimental data into account. The posterior distribution cannot be found analytically, and so we use Markov Chain Monte Carlo (MCMC) methods \citep[e.g.,][]{Gilks96} to  obtain an approximation.
We describe the case where only the $n$ and $m$ values, the number of eggs that reached maturity, have been recorded. The simpler situation where $N_i$ is observed is a special case and follows immediately.

We introduce prior distributions for all unknowns. We assume the number of eggs laid in each clutch follows a Poisson distribution with mean $\lambda$
\begin{equation}\label{eqn:Poisson}
 N_i \sim Po(\lambda) \mbox{ for } i=1,\ldots, C
\end{equation}
and for the fixed parameters we assume  that
\begin{equation}\label{eqn:priors}
\begin{array}{ll} 
p \sim U[0,1] & \qquad \psi \sim N(0,\sigma^2)\\
\lambda \sim \Gamma(a, b) & \qquad d \sim \operatorname{Beta}(a',b').\\
\end{array}
\end{equation}
The distribution of $p$, $\lambda$ and $d$ are conjugate to the likelihood, allowing a Gibbs sampler to be used. Informative prior distributions are usually available for $\lambda $ and $d$, as scientists often have information about mortality rates and average clutch sizes for the species of interest, although simulation suggests that the quantities of interest (the Bayes factor and the posterior of $\psi$), are robust to the choice of priors for $\lambda$ and $d$.
The key parameter is the dispersion parameter $\psi$, which we assign a zero mean normal distribution, so that under- and over-dispersion are equally likely {\it a priori}. 
We use an uninformative prior distribution for $p$, so that the posterior distribution is determined solely by the data. 

To sample from the posterior distribution $\pi(\theta | D)$, we use a Metropolis-Hastings within Gibbs sampler \citep{Metropolis53, Geman84}. We introduce vectors of unobserved $N_i$ and $M_i$ values, denoted $\bf{N}$ and $\bf{M}$, as auxiliary variables, and sample across the chain 
$\pi (\theta, {\bf N}, {\bf M}| D)$, which is a $(4+2C)$ dimensional Markov chain. The distribution of interest, $\pi(\psi|D)$, is then found by taking the marginal distribution of $\psi$. 
Details of the MCMC algorithm used are provided in the supplementary material, and the algorithm is implemented in the accompanying {\tt precision} R package for each of the three models.

\subsection{Bayes factor estimation}

To calculate the Bayes factors (Equation \ref{eqn:BF}) we must first  calculate the evidence for each model
\[\pi({\bf n}, {\bf m}) = \int \pi({\bf n}, {\bf m}|\theta) \pi(\theta) {\rm d} \theta,
\]
where ${\bf n}$ and ${\bf m}$ are the vectors of the observed $n_i$ and $m_i$ values, which is analytically intractable for  the models  considered. We use the approach described in \citet{Chib95} and \citet{ChibJeliazkov01} to estimate the evidence for each model, which relies upon the identity
\begin{equation}\label{eqn:Chib}
\pi(\bn,\bm) = \frac{\pi(\bn,\bm|\theta^*)\pi(\theta^*)}{\pi(\theta^*|\bn,\bm)}.
\end{equation}
Calculation of both the numerator and denominator is challenging, but can be done with additional samples from an MCMC sampler. The derivation and details of the algorithm are technical, and are presented in the supplementary material. 
An implementation of  these algorithms is available as  the {\tt precision} R package,  available on github. The next section demonstrates the power of our approach.

\section{Results}\label{sect:results}

We illustrate our  approach using data on four species of wasp: 
The strength of evidence for under-dispersion from secondary sex ratio data in these species varies from  weak ({\it Colpoclypeus florus}) to  overwhelming ({\it Metapycus luteolus}), and the mortality rate varies from low ({\it Goniozus legneri}) to high ({\it C. florus}). We also present the results from a  simulation study which conclusively demonstrates the increased power of our approach.

The Bayesian approach requires  prior distributions for all unknown parameters. Simulation studies have shown that the model and data are strongly informative about $p$ and $\psi$, so that any information in the prior distribution  is overwhelmed by the information in the data. In all our analyses we give $p$ an  uninformative prior distribution uniform on $[0,1]$ and $\psi$  a vague  prior distribution for both the double and multiplicative binomial models:
\begin{equation}\label{eqn:ppsipriors}
p\sim U[0,1],\qquad  \psi \sim N(0,1).
\end{equation}
The prior for $\psi$ can be justified by examining the degree of under-dispersion for various levels of $\psi$. If $M \sim \operatorname{DoubleBinom}(n=10, p=0.1, \psi)$, then $\BP(M=1) = 0.38$ if $\psi=0$ (the binomial case), whereas for $\psi=3$, $\BP(M=1) = 0.85$, indicating  strong under-dispersion. The Bayes factors are robust to the choice of priors for $\lambda, d $ and $p$ (the parameters shared across models), but  unsurprisingly, are sensitive to the prior for $\psi$. More diffuse priors for $\psi$ tend to reduce the evidence for under-dispersion due to an Occam's razor type effect, but for realistic priors for $\psi$, the conclusion does not usually change significantly (see supplementary material). Fortunately, the posterior distribution for $\psi$ is robust to the choice of prior for $\psi$, and so this can also be used to indicate whether the data are under-dispersed.

 The data typically contain only limited information  about the parameters $\lambda$ and $d$, but with the  two posterior distributions  strongly correlated, as large average clutch size and high mortality, or small average clutch size and lower mortality rate, leads to similar datasets. Prior information about $\lambda$ and $d$ is often available, which we can use to choose prior distributions for these two parameters on a species by species basis. Experimentation has shown that the Bayes factor and the posterior distribution of $p$ and $\psi$ (the primary parameter of interest) are robust to these choices.

\subsection{{\it Goniozus legneri:} Large dataset, low mortality}\label{sect:Glegneri}

We begin by considering  data on {\it G. legneri}, a gregarious parasitoid wasp in which offspring groups are produced by single mothers and sex ratios are female biased due to local mate competition. \citet{Khidr_etal2013} provide both a primary dataset, consisting of pre-mortality counts on 47 clutches obtained using DNA microsatellite markers to identify the sex of eggs,  and a secondary dataset containing
post-mortality counts of male and female adults in 113 clutches. Both the Meelis and James tests lead to   rejection of the null hypothesis of binomial  sex allocation (Table \ref{table:Results}) with $p$-values of 0.0041 and 0.0027 respectively for the secondary data. Furthermore, we find $R=0.572$, which when combined with the  negative value of $U$ in the two tests ($U=-2.38$ and $U=-1.98$ for Meelis and James respectively), lead us to conclude, in common with previous studies \citep{Hardy_etal1998, Khidr_etal2013}, that {\it G. legneri} has under-dispersed sex ratios.

\begin{sidewaystable}
\centering
\begin{tabular}{lll llllll}
\toprule
&& \multicolumn{7}{c}{{\bf Species}}\\
\midrule
 && \multicolumn{2}{l}{\it G. legneri} & {\it G. thailandensis} & \multicolumn{2}{l}{\it C. florus} & \multicolumn{2}{l}{\it M. luteolus}\\
\midrule
Proceedure & Instance& Primary & Secondary & Secondary & Primary & Secondary& Primary & Secondary\\
&&\multicolumn{1}{l}{Value}&\multicolumn{1}{l}{Value}&\multicolumn{1}{l}{Value}&\multicolumn{1}{l}{Value}&\multicolumn{1}{l}{Value}&\multicolumn{1}{l}{Value}&\multicolumn{1}{l}{Value}\\
\midrule
James & U & -1.98 &  -3.00 & -2.01 & -0.89 & 2.7& -6.7& -7.8\\
		 &p & 0.047 &0.0027 & 0.045 & 0.37 & 0.0068&$1.5\times 10^{-11}$&$7.0\times 10^{-15}$\\
		\cmidrule(lr){2-9} 
		 Meelis &U & -2.38& -2.87 & -0.73 & -3.24 & -0.97&-7.9 & $-7.4$\\
		 &p &0.017  & 0.0041 & 0.46 & 0.0012 & 0.33&$2.6\times 10^{-15}$&$1.3\times 10^{-13}$\\
		\cmidrule(lr){2-9}
		$R$& &0.44 &  0.57 & 0.68 & 0.13 & 0.75& 0.093  & 0.44\\		
		$s^2$ & &0.57 & 0.61 & 0.74 & 0.51 & 1.18& 0.20 & 0.58\\
\midrule
\addlinespace[0.5em]
BF& double:binomial & 45.1 &213.6 & 5.65 & 3830 & 0.27& $1.1 \times 10^{29}$ &$9.8\times 10^{23}$\\
& multiplicative:binomial & 9430&31.3& 0.54 & 0.36 & 0.36 &  $7.0 \times 10^{5}$ & $2.0\times 10^{6}$\\
& double:multiplicative & 0.0048&6.8 & 10.5 & 10600 & 0.74&$1.6\times 10^{23}$ & $5.0\times 10^{17}$\\
\cmidrule(lr){2-9}
 Posterior & binomial & 0.00010 &0.004 & 0.14 & 0.00026 & 0.61&0.000 & 0.000\\
probability & multiplicative & 0.995&0.127& 0.074 & 0.000094 & 0.22&0.000&0.000\\
 &double & 0.0048&0.869 & 0.74 & 0.9996 & 0.16& 1.000& 1.000\\
\bottomrule 
\end{tabular}
\caption{Analysis of four wasp datasets: {\it G. legneri} primary ($C=47$) and secondary ($C=113$) datasets  \citep{Khidr_etal2013}; {\it G. thailandensis} secondary dataset ($C=60$)   \citep{Witethom_etal1994}; {\it C. florus} primary ($C=55$) and secondary datasets ($C=53$)  \citep{Dijkstra1985, Hardy_etal1998};
 {\it M. luteolus} primary ($C=127$) and secondary ($C=371$) datasets \citep{Kapranas_etal2011}.  
All values estimated using $10^6$ MCMC iterations.}
\label{table:Results}
\end{sidewaystable}

\begin{figure}
\centering
\includegraphics[width=\textwidth]{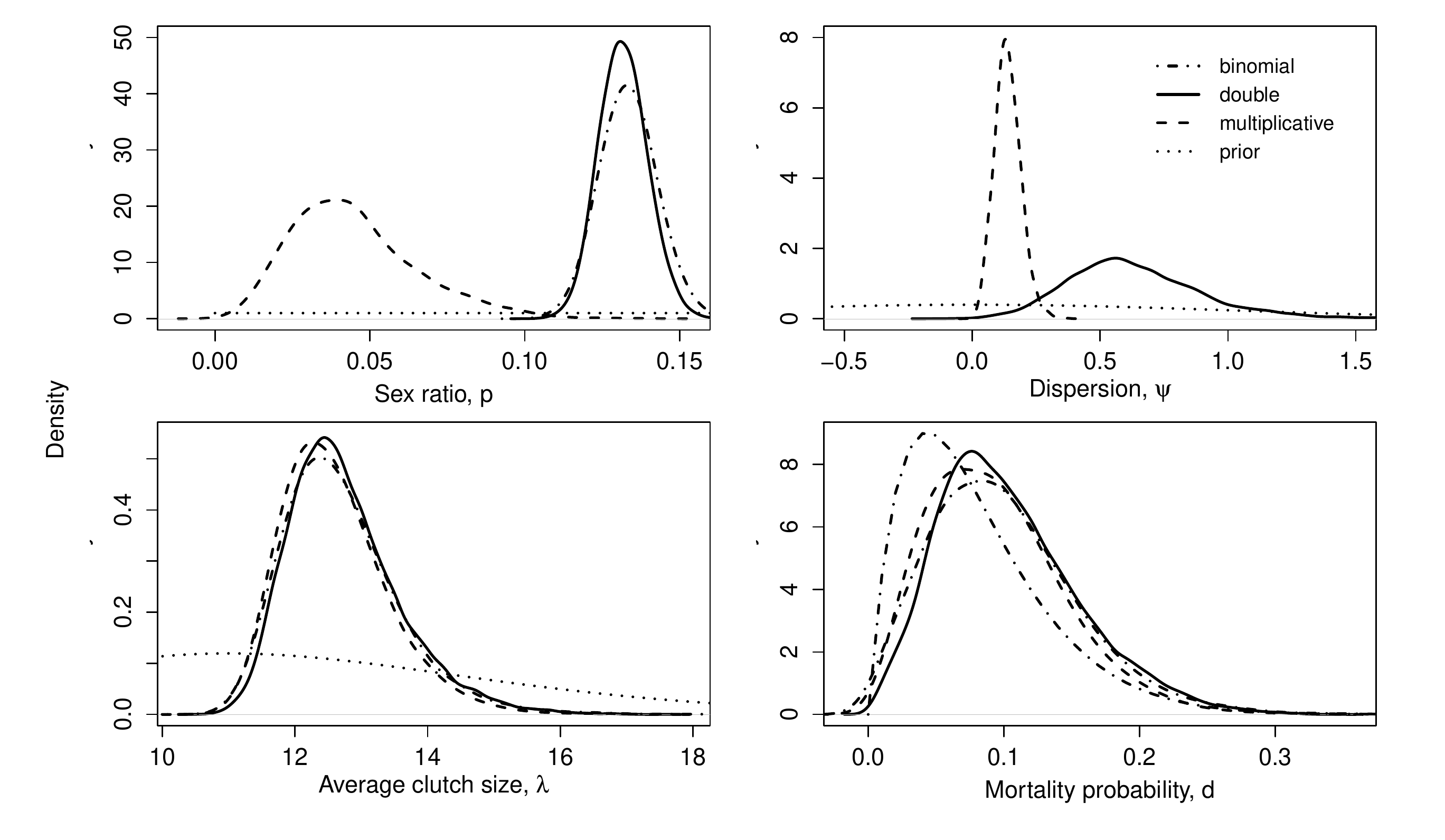}
\caption{Posterior distributions from the analysis of   {\it G. legneri} secondary data \citep{Khidr_etal2013},  obtained using $5\times 10^5$ MCMC iterations. For each parameter, the prior and posterior distribution are shown for the three alternative models of sex allocation. Note that the binomial model does not have a dispersion parameter ($\psi$)  and that the interpretation of $p$ and $\psi$ is different in each model.}
\label{fig:Glegneri}
\end{figure}

\citet{Khidr_etal2013} reported that the proportion of offspring that died before maturity was 7.6\%, which agrees with  previous {\it G. legneri} mortality estimates \citep[5-12\%,][]{Hardy_etal1998}. We incorporate this information into the analysis through the use of prior distributions
\[d\sim \operatorname{Beta}(2,23)\qquad \lambda \sim \operatorname{Gamma}(12,1).\]
The prior mean for $d$ is thus $2/(23+2)=8\%$, with values in the range 0-20\% all supported {\it a priori} (Figure \ref{fig:Glegneri}). The prior for $\lambda$ was based on an observed secondary clutch size of 11, and the mortality rate of 7.6\%, suggesting a prior mean for $\lambda$ of approximately 12. The  $\operatorname{Gamma}(12,1)$ prior distribution has a prior mean of $12/1$, and supports prior $\lambda$ values in a range between 11 and 14 (Figure \ref{fig:Glegneri}).

Figure \ref{fig:Glegneri} shows the posterior distributions of the four parameters for the secondary dataset. Interest lies primarily in the dispersion parameter $\psi$, with $\psi>0$ indicating under-dispersion and $\psi<0$ over-dispersion. We cannot estimate $\psi$ precisely as there is a finite quantity of data, but the posterior distributions show  the range of $\psi$ values we believe could feasibly have led to the observed data. The posterior distribution for $\psi$ for both the double and multiplicative models, suggests that only positive values of $\psi$ are consistent with the data. 
Equi-tailed 95\% credibility intervals for $\psi$ are $[0.047, 0.248]$ for the multiplicative binomial model,
and $[0.196, 1.28]$ for the double binomial model, neither of which overlap with $0$, leading us to conclude that {\it G. legneri} has under-dispersed sex allocation. 

The Bayes factor (BF) estimates for {\it G. legneri} are reported in Table \ref{table:Results}. We find that the double binomial model is best supported, with a BF of 213.6 in favour of the double binomial over the binomial model, which Jeffreys' scale interprets as decisive evidence. There is also very strong evidence in favour of the multiplicative model over the binomial ($BF=31.3$), and substantial evidence to suggest the double binomial is better supported than the multiplicative binomial model ($BF=6.8$). If we are prepared to assign all three models equal prior probability, then the posterior probability that the binomial model is the true model is 0.004, compared to 0.869 for the double binomial model, and 0.127 for the multiplicative binomial model.

\begin{figure}
\centering
\includegraphics[width=\textwidth]{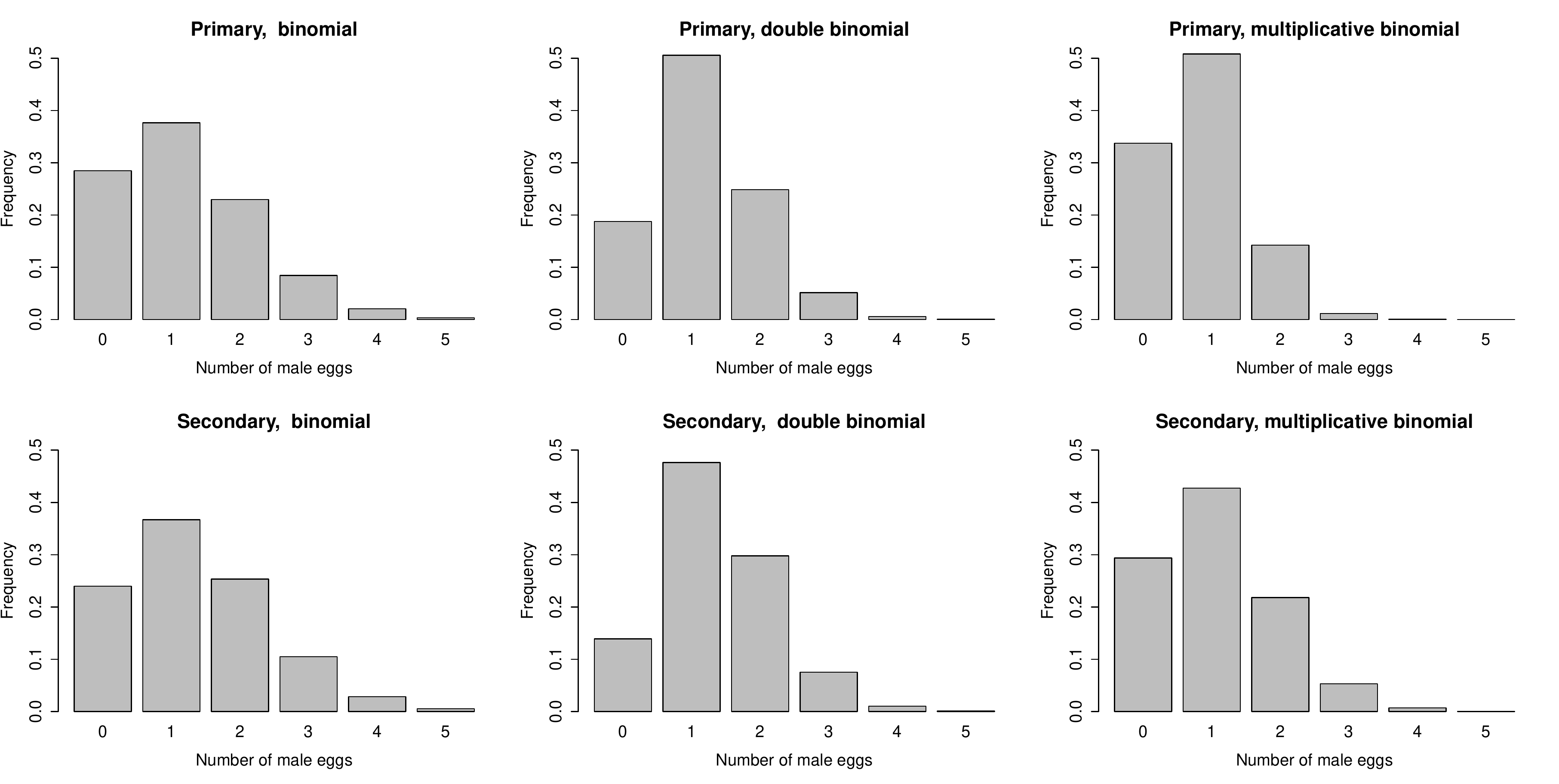}
\caption{Posterior predictive distributions of the number of male eggs (pre-mortality) in a clutch of 10 eggs for the six different scenarios, namely  the primary and secondary data for the three different models of sex allocation. The multiplicative binomial distribution gives the best fit to the primary data, and the double binomial distribution  best fits  the  secondary data.}  
\label{fig:PostPred}
\end{figure}

For this dataset, the signal from the data is strong ($C=113$ is a reasonably large sample size), and consequently all the procedures  give unambiguous conclusions. However, it is informative to note the difference between the two approaches: the Meelis test strongly rejects $H_0$, but does not indicate the size of the effect (the R value does indicate the size of the effect, but  is  unreliable without a measure of uncertainty). The $p$-value does not give the probability that $H_0$ is true and should not be interpreted as such. Meanwhile, the Bayesian procedure estimates the probability that $H_0$ is true, and the posterior distribution  for $\psi$ gives the effect size after having accounted for mortality, along with a measure of the uncertainty in the estimate of $\psi$.
For {\it G. legneri}, \citet{Khidr_etal2013} also provide a primary dataset which we can analyse without  modelling mortality (Table \ref{table:Results}). The conclusion is the same as for the secondary data, again  with strong evidence of under-dispersion. One difference between the primary and secondary analyses is that for the primary data, the multiplicative binomial model is preferred, whereas for the secondary data, the double binomial model is preferred. We believe this is due to differences between the shape of the two distributions. Figure \ref{fig:PostPred} shows the posterior predictive distribution for the number of male eggs laid (in a clutch of 10 eggs)  for the six different scenarios (three models on both the primary and secondary data). We can see that for a given sex allocation model, the posterior predictions for the primary and secondary data are similar, and that the double and multiplicative distributions both give more concentrated (more precise) predictions than the binomial model. We can also see the difference between the shape of the double and multiplicative distributions, with the multiplicative distribution predicting more clutches with no males than the double binomial. The switch between preferred model for the secondary and primary datasets does not change our conclusion that there is strong evidence of  under-dispersion.

Finally, note that the data and model are strongly informative about $p$ and $\psi$, with the posterior and prior values being markedly different, whereas the posterior value for $\lambda$ and $d$ are close to the prior distribution. Experimentation (see the supplementary material) has shown that the posterior distributions of $\lambda$ and $d$ are sensitive to their prior distribution, but that the posterior of $p$ and $\psi$ are not sensitive to these choices.

\subsection{{\it Goniozus thailandensis:} small dataset, medium mortality} \label{sect:Gthail}

Now we consider a dataset on the parasitoid species {\it Goniozus thailandensis} collected by \citet{Witethom_etal1994}. This species has a broadly similar biology to {\it G. legneri} and has previously been analysed for sex ratio variance by \citet{Hardy_etal1998}.  The developmental mortality rate, 22\%, is higher than for {\it G. legneri} and the dataset is small, thus presenting a more challenging, and possibly more typical, case for analysis. 
Classical analysis of these data was inconclusive: the Meelis test gave $U=-0.73$ with a $p$-value of 0.23 and $R=0.68$,
which suggests under-dispersion,
 but with insufficient evidence to reject $H_0$ at the 5\% significance level. In Section \ref{sect:Current} we demonstrated that the Meelis test will lack power on this dataset, as there are only $C=60$ observations and the probability of developmental mortality is moderate. This leaves us uncertain as to whether this result is due to the  limited sample size, the relatively high mortality rate or to sex allocation actually being binomially distributed. The Meelis test only informs us that we cannot reject the null hypothesis due to insufficient evidence; it does not allow us to say that the species has binomially distributed sex allocation.

\begin{figure}
\centering
\includegraphics[width=\textwidth]{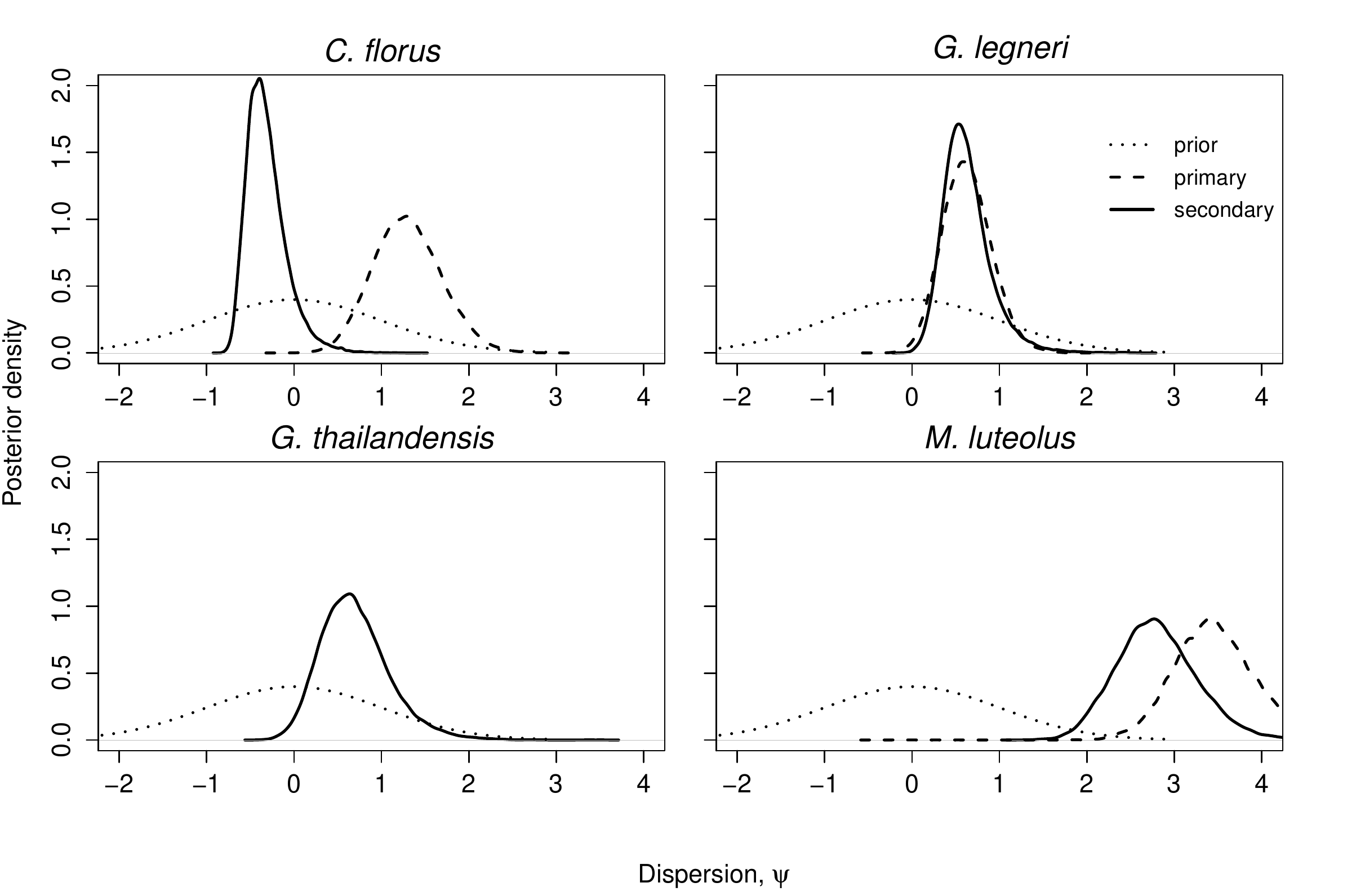}
\caption{The marginal posterior distribution for $\psi$ for the double binomial model for all four species. The results were obtained using $5\times 10^5$ MCMC iterations. The posterior distributions for $p$, $d$ and $\lambda$ are not shown.}  
\label{fig:posteriors}
\end{figure}

Carrying out the Bayesian analysis, using the prior distributions 
\[d\sim \operatorname{Beta}(5,20)\qquad \lambda \sim \operatorname{Gamma}(9,1),\]
(consistent with the observed average clutch size and the mortality rate of 22\%) we find the posterior distribution for $\psi$ shown in the bottom left panel of Figure \ref{fig:posteriors} and the Bayes factors given in Table \ref{table:Results}. The Bayes factors suggest that there is substantial evidence in favour of the double binomial model over the other two models, and the posterior for $\psi$ shows that  under-dispersion is the best explanation of the data (the equi-tailed 95\% credibility interval for $\psi$ is $[0.0415,  
1.61]$). The posterior distribution does contain a small amount of support for a zero or negative value of $\psi$ (binomiality, or over-dispersion), showing that while this can not conclusively be ruled out, it is unlikely.
Assuming equal prior probability for each model, there is a posterior probability of 0.79 that the double binomial model is the true model, and 0.14 that the binomial model ($H_0$) is true. While this is not conclusive evidence, it has allowed us to state that the data suggest under-dispersion over binomial sex allocation with posterior odds of more than 5 to 1. The posterior for $\psi$ allows us to see the range of possible under-dispersion strengths that are consistent with the data. In comparison, the classical approach only allows us to conclude that there is insignificant evidence to reject $H_0$.

\subsection{{\it Colpoclypeus  florus:} medium dataset, high mortality}\label{sect:Cflorus}

Primary and secondary data on {\it Colpoclypeus florus} are available from a study by \citet{Dijkstra1985} analysed by \citet{Hardy_etal1998}. {\it C. florus} is a gregarious parasitoid with female biased sex ratios and is the only known member of its genus. 
The mortality rate was reported to be 57\%, which when combined with the average clutch size of 7.4 motivated the prior distributions
\[d\sim \operatorname{Beta}(11,10)\qquad \lambda \sim \operatorname{Gamma}(16,1).\]
The results of the   analysis of this data are shown in Table \ref{table:Results}. These illustrate the tendency  of mortality to make data appear less under-dispersed, possibly even over-dispersed. The primary data clearly show that the species has under-dispersed sex allocation, with the Meelis test and Bayes factors agreeing that there is very strong evidence in favour of under-dispersion. Whereas for the secondary data, the Meelis test fails to reject the null hypothesis, and the Bayes factors suggest that the binomial model is the best supported (posterior probability of 0.61, compared to 0.16+0.22=0.38 for the two non-binomial models). The 95\% credibility interval for $\psi$ is  
$[-0.063, 0.019]$ for the multiplicative model, and $[-0.65, 0.24]$ for the double binomial model, both of which contain $0$, showing that the data could be either under- or over-dispersed. The marginal posterior for $\psi$ in Figure \ref{fig:posteriors}, shows how the primary data strongly suggest under-dispersion, but that the secondary data (after mortality) suggest over-dispersion, although there is still some support for under-dispersion. 
	While the Meelis test can only lead us to conclude that there is no evidence to reject the hypothesis of binomial sex allocation, the Bayesian test can quantify that evidence and give a posterior probability that indicates that the hypothesis of binomial sex ratios is approximately twice as likely as the hypothesis of non-binomial sex allocation.

\subsection{{\it Metaphycus luteolus:} large dataset, high mortality}

A large secondary  dataset on {\it M. luteolus} was presented in \citet{Kapranas_etal2011}. This species is a facultatively gregarious parasitoid which lays eggs inside hosts. Developing offspring may compete within the host, be attacked by the host immune responses, or die of other causes, and the overall mortality rate is approximately 40\%. The secondary sex ratio is female biased. Using  prior distributions
\[d\sim \operatorname{Beta}(6,10)\qquad \lambda \sim \operatorname{Gamma}(4,1)\]
we obtained the results presented in Table \ref{table:Results} and Figure \ref{fig:posteriors}. Due to the large sample sizes, and the effect size, all procedures give overwhelming evidence that the data are under-dispersed. By  selecting only those  clutches that did not experience any mortality, we can obtain an approximation of  a primary dataset  \citep[this approach is discussed in][]{Khidr_etal2013}. Analysis of this dataset again demonstrates the tendency of mortality to make data appear less under-dispersed.

\subsection{Simulation study}
We now show that by modelling mortality, we have increased our ability to detect under-dispersion. We analyse the performance of the Meelis test and the  Bayes factor approach, using a simulation study in which we apply both  procedures to  synthetic datasets. The computational expense of the Bayesian approach (typically it takes 2-5 hours of computer time to analyse a single dataset), limited the study  to 100 synthetic datasets, but this is sufficient to conclusively demonstrate an improved ability to find evidence against  $H_0$, i.e., statistical power.

The synthetic datasets were simulated to each contain $50$ clutches using a mortality rate of 30\%,  moderate values of $C$ and $d$ 
The model defined by Equations (\ref{eqn:multbinom}), (\ref{eqn:mort1}) and (\ref{eqn:Poisson}), with $\lambda=10, p=0.1$, and $\psi =0.3$, was used to simulate the datasets, giving a moderate level of under-dispersion comparable to {\it G. legneri}.

\begin{figure}
\centering
\includegraphics[width=\textwidth]{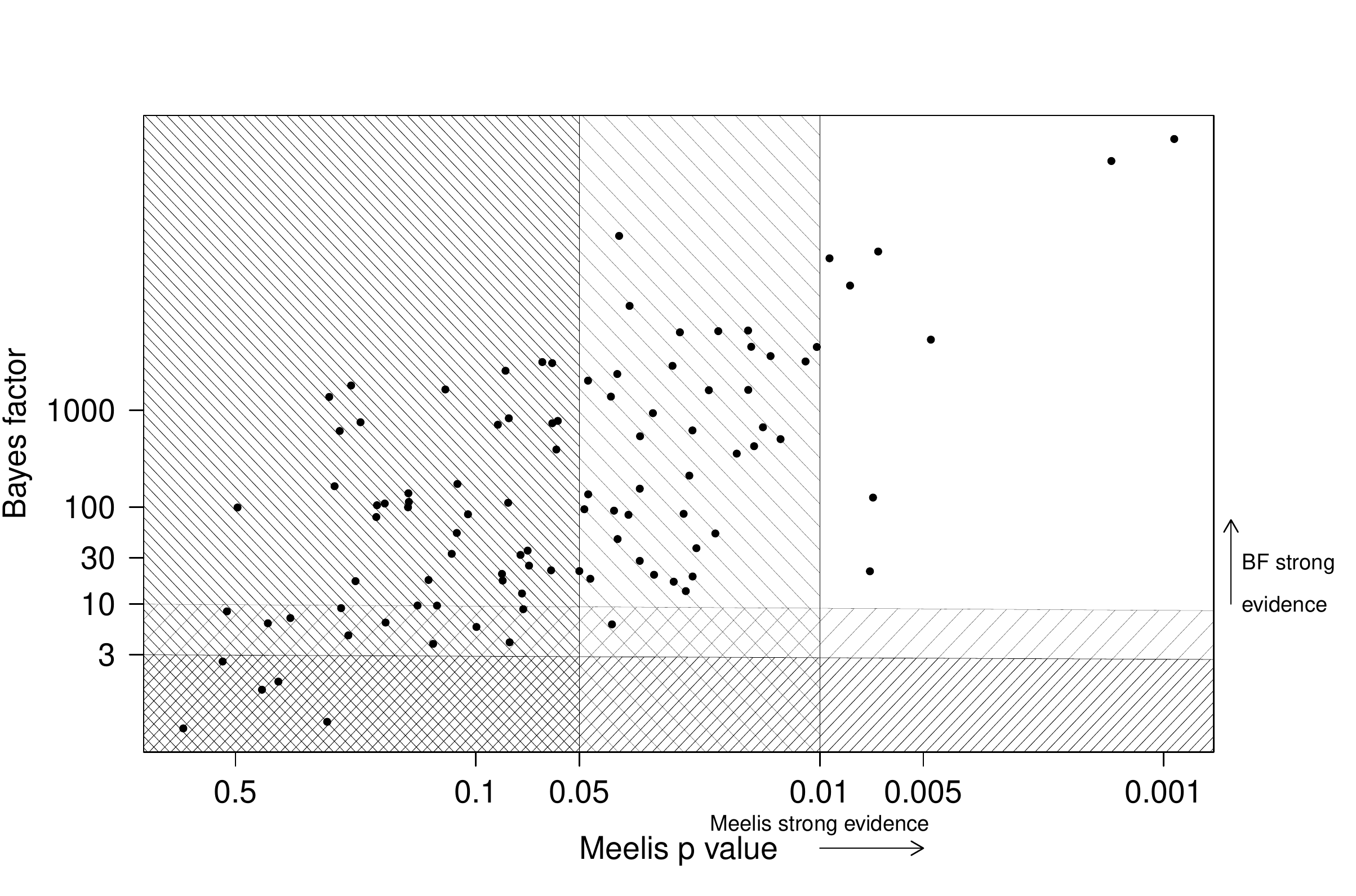}
\caption{
Results of the simulation study. Each point represents the $p$-value and Bayes factor (BF) for a simulated dataset. The shaded regions indicate $p$-value or Bayes factor ranges for which we would conclude there was either no, or weak evidence against $H_0$. The horizontal  regions (with `forwardslash' shading) indicates the Bayes factor is either less than 10 (threshold for strong evidence against $H_0$), or less than 3 (threshold for substantial evidence). The vertical  regions (`backslash' shading) indicate $p$-values of less than 0.01  or 0.05. Note that the x-axis is reversed.}
\label{fig:SimStudy}
\end{figure}

The results of the simulation study are summarised in Figure \ref{fig:SimStudy} and Table \ref{table:SimStudy}. For each dataset we have plotted the logarithm of the estimated Bayes factor between the multiplicative and binomial models,  against the logarithm of the $p$-value from the Meelis test. The shading shows  regions in which one or both of the procedures  failed to detect strong evidence of under-dispersion, either because the $p$-value is greater than $0.05$ (or $0.01$), and/or because the Bayes factor is less than 3 (or  10).  Table \ref{table:SimStudy}  summarises each procedure by the percentage of datasets which led to  Bayes factors or $p$-values in a specified range.

\begin{table}
\centering
  \begin{tabular}{l c c c c c}
\toprule
Strength of\\
evidence:
 &\multirow{-2}{*}{insubstantial} &  \multirow{-2}{*}{substantial}& \multirow{-2}{*}{strong} & \multirow{-2}{*}{very strong}& \multirow{-2}{*}{decisive}\\
 \midrule
Meelis $p$-\\
value range: & \multirow{-2}{*}{$>0.1$} & \multirow{-2}{*}{0.05-- 0.1} & \multirow{-2}{*}{0.01--0.05} & \multirow{-2}{*}{0.001-- 0.01} & \multirow{-2}{*}{$<0.001$}\\
\% in range: & 33 &20& 39 & 7 & 1\\
\cmidrule(lr){2-6}
BF range: & 0 -- 3 & 3 -- 10 & 10--30& 30-100 & $>100$\\
\% in range:& 5 &13& 15& 15& 52\\
\bottomrule
 \end{tabular}
\caption{Simulation study results: 100 synthetic datasets, all with moderate levels of under-dispersion ($\psi=0.3$ in the multiplicative model) and mortality (30\%), were analysed and grouped into categories indicating various levels of strength of evidence against $H_0$. The Bayesian approach can be seen to substantially outperform the Meelis test.}
\label{table:SimStudy}
\end{table}

These  results clearly demonstrate the improved power of the Bayesian procedure. For example, in more than half of the simulated datasets, the Meelis test returned a $p$-value greater than 0.05, which would indicate that there was insufficient evidence to reject the null hypothesis of binomial sex allocation. In contrast, 95\% of the datasets provided at least substantial evidence against binomial sex ratios according to the Bayesian approach, and 
over half (52\%) of the datasets provided decisive evidence ($BF>100$). Furthermore, Figure \ref{fig:SimStudy} illustrates that every time the Bayesian test failed to detect under-dispersion, the Meelis test also failed, whereas  there were 36   datasets  where the Bayesian test indicated strong evidence ($BF>10$) against $H_0$, but where the Meelis test failed (at the 5\% level).

In order to confirm that this increased power is not due to a corresponding increase in the type I error rate (i.e., falsely rejecting $H_0$), a second simulation study was performed  analysing synthetic datasets generated from the binomial model. For 200 simulated datasets, the Meelis test rejected $H_0$ (at $\alpha= 0.05$) in 3\% of cases (i.e., it had approximately the assumed error rate). The Bayes factor gave $\BP(H_0 | D) \leq 0.05$ (i.e., strong evidence against $H_0$) in 6\% of cases, showing that the increased power of the Bayesian approach is not due to an inflated type I error. 
The posterior distributions for $\psi$ (available in the supplementary information),   ruled out $\psi=0$ in only one of the 200 simulated datasets.

\section{Conclusions}

We have shown that  the current approaches used to detect under- or over-dispersion in sex allocation lack power when the sample size is small or the mortality rate is moderate to large. Both are common situations in empirical studies. For example, the Meelis test will usually fail to reject the null hypothesis under these conditions even when sex allocation is strongly non-binomial. We have introduced a new approach to detecting under- or over-dispersion that has much greater power for detecting departures from binomial allocation. The approach gains its power by explicitly modelling mortality, so that the test takes into account that the patterns in the data have occurred through a combination of sex allocation and mortality. The method can be extended further to include non-binomial distributions of mortality \citep[e.g.,][]{Hardy_etal1998, Kapranas_etal2011}. Furthermore, using a Bayesian approach to model selection and parameter estimation increases our descriptive ability: the posterior distribution of the dispersion parameter $\psi$ allows both the size of the effect and the range of possible effects that are consistent with the data to be identified. Using Bayes factors allows us to give the posterior probability that the data derive from a species that has binomially distributed sex allocation, as opposed to $p$-values, which although commonly interpreted as probabilities, should not be \citep{Goodman08}. In situations where the evidence is conclusively in favour of one hypothesis, our test generates the same conclusion as current approaches (but with improved descriptive ability). However, when the evidence is weaker, the additional information provided by the Bayesian approach can  allow us to make useful inferences, even if these cannot be conclusive.


\section{Coda}
The software implementing this approach has been written in R \citep{R} and is freely available ({\tt https://github.com/rich-d-wilkinson/precision}) as the {\tt precision} R package on github. Details of how to use and install the package are given in the package vignette and in the supplementary material.
There are many possible extensions to this approach, primarily through changes and improvements to the model. For example, the binomial mortality model is relatively simple and other more complex models (such as over-dispersion) are possible. These extensions are straightforward to make within the Bayesian testing framework. 

The data used in this paper are all available within the {\tt precision} R package (see the package vignette). These datasets, as well as additional data on the  sexual compositions of offspring groups, are available from several previous publications. Secondary sex ratio datasets can be found in \citet{Morgan1994, Hardy1995, NagelkirkeSabelis98, Mackauer_etal2002, Kapranas_etal2008, Kapranas2009} and \citet{Khidr_etal2013}. Primary sex ratios are more difficult to evaluate, but datasets are available in \citet{Dijkstra1985, Aviles2000}, and \citet{Khidr_etal2013}. 


\medskip

\noindent {\bf Acknowledgements:} We thank S. K. Khidr, B. Witethom and L. J. Dijkstra for help with data. 
We thank Andrew Wood for useful advice on the statistical approach taken, and R. F. Green and three anonymous referees for constructive suggestions.
Apostolos Kapranas was funded by a Marie Curie Fellowship (FP7-PEOPLE-IEF-273431). 
\bibliographystyle{elsarticle-harv}
\bibliography{Revised_JTP}



\end{document}